\documentclass[twocolumn,aip,cha,
amsmath,amssymb]{revtex4}
\usepackage{bm}
\usepackage[colorlinks=true,linkcolor=blue]{hyperref}
\usepackage{amsmath}
\usepackage{amssymb}
\usepackage{amsthm}
\usepackage{amsfonts}
\usepackage{enumerate}
\usepackage{latexsym}
\usepackage{ifpdf}
\usepackage{graphicx}
\expandafter\ifx\csname package@font\endcsname\relax\else
 \expandafter\expandafter
 \expandafter\usepackage
 \expandafter\expandafter
 \expandafter{\csname package@font\endcsname}%
 \fi
\begin{document}
\title{Local Electronic and Magnetic Studies of an Artificial La$_2$FeCrO$_6$ Double Perovskite}
\author{Benjamin Gray$^1$, Ho Nyung Lee$^2$, Jian Liu$^1$, J. Chakhalian$^1$, and J. W. Freeland$^3$}
\affiliation{$^1$Department of Physics, University of Arkansas, Fayetteville, AR 72701}
\affiliation{$^2$Materials Science and Technology Division, Oak Ridge National Laboratory, Oak Ridge, TN 37831}
\affiliation{$^3$Advanced Photon Source, Argonne National Laboratory, Argonne, IL 60439}

\begin{abstract}
Through the utilization of element-resolved polarized x-ray probes, the electronic and magnetic state of an artificial La$_2$FeCrO$_6$ double perovskite were explored. Applying unit-cell level control of thin film growth on SrTiO$_3$(111), the rock salt double perovskite structure can be created for this system, which does not have an ordered perovskite phase in the bulk. We find that the Fe and Cr are in the proper 3+ valence state, but, contrary to previous studies, the element-resolved magnetic studies find the moments in field are small and show no evidence of a sizable magnetic moment in the remanent state.
\end{abstract}

\date{\today}
\maketitle


The double-perovskite crystal structure, A$_2$BB$^{\prime}$O$_6$ provides a rich framework for engineering materials with unrealized electronic and magnetic properties \cite{Pickett1}. By rationally designing superlattices, these features can be exploited to design artificial magnets. Specifically, the 180$^o$ superexchange between transition metal (TM) ions which occupy the B and B$^{\prime}$ sublattices can be goverened through the choice of TM, growth direction, stacking periodicity, etc. \cite{Ueda, Ueda2, Ueda3}.

La$_2$FeCrO$_6$ is a model system in theoretical discourses over the Goodenough-Kanamori rules. \cite{Kanamori}. However, numerous experimental attempts at conventional chemical bulk synthesis have failed to produce an ordered phase of La$_2$FeCrO$_6$ \cite{Wold, Belayachi, Belov}. Ueda et al. employed multi-target pulsed laser deposition to overcome this difficulty \cite{Ueda, Ueda2, Ueda3}. By stacking alternating monolayers of stoichiometric LaCrO$_3$ and LaFeO$_3$ on a SrTiO$_3$ (STO) substrate orientated along the (111) plane (see Fig. 1), an artificial double perovskite with rock salt B-site ordering was fabricated. For this system, an average magnetization of 3 $\mu$$_B$ per transition metal ion was measured with a superconducting quantum interference device (SQUID) magnetometer and an undifferentiated B-site charge state [Cr$^{3+}$(3d$^3$), Fe$^{3+}$(3d$^5$)] with ferromagnetic coupling between Cr and Fe was deduced, which agrees with the Goodenough-Kanamori rules \cite{Ueda, Goodenough, Kanamori}. However, a post publishing technical comment showed that the actual magnetization was 0.009 $\mu$$_B$ per site at low field\cite{Meijer}, which raises questions concerning the nature of the ordered magnetic state. 

In this letter, we report on the electronic and magnetic structure of an artificial La$_2$FeCrO$_6$ double perovskite that is grown as a single unit cell LaFeO$_3$/LaCrO$_3$ superlattice (SL) on STO (111) substrates. X-ray absorption spectroscopy (XAS) shows that both Cr and Fe are trivalent in an O$_h$ ligand field symmetry and high spin state. At low temperatures and large magnetic fields, X-ray Magnetic Circular Dichroism (XMCD) shows that the magnetic moments of Fe and Cr are coupled ferromagnetically. However, the moments are only a tenth of the expected values for a saturated ferromagnetic state and the signal disappears in zero magnetic field, which are both suggestive of an anti-ferromagnetic state. 

Using the element specific techniques of XAS and XMCD,  the local electronic and magnetic properties of LaCrO$_3$ and LaFeO$_3$ superlattices (SL) were investigated. XAS/XMCD is defined as the sum/difference between the absorption spectra for left and right-polarized x-rays. The sample was mounted inside a superconducting magnet system with incidence angle of 10$^{\circ}$ from the sample plane (80$^{\circ}$ from the growth axis). Since the beam and magnetic field are parallel, this aligns the field primarily in-plane. The polarization of the x-ray beam is switched between left- and right-circularly-polarized light at each energy interval in order to measure the photon helicity and magnetization parallel (I$^+$) and antiparallel (I$^-$). XAS (I$^+$ + I$^-$) stems from the electronic character of a particular element in its environment, while XMCD's (I$^+$ - I$^-$) magnetic origins probe the element specific net magnetization. Due to XMCD's monolayer sensitivity and penetration depth of several hundred angstroms, it is a prime candidate for exploring nanoscale magnetism at interfaces. In the past, this technique was successfully used to study ultra-thin films of manganites, cuprates, cobaltites, etc. \cite{ChakhalianSCI, ChakhalianNAT}.

\begin{figure}[t]\vspace{-0pt}
\includegraphics[width=8.5cm]{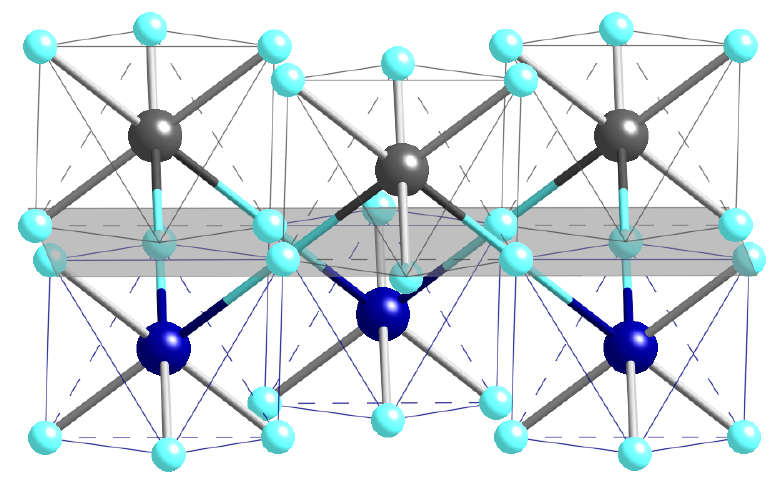}
\caption{\label{Growth111} Model of interface between LaFeO$_3$ and LaCrO$_3$ monolayers grown in the [111] direction. Superexchange path is indicated by multi-colored bonds. Layer above plane represents LaFeO$_3$. Layer below plane represents LaCrO$_3$.}
\end{figure}

The single unit cell LaFeO$_3$/LaCrO$_3$ SL was fabricated on commercially available STO (111) substrates by pulsed laser deposition with in-situ monitoring by high-pressure RHEED \cite{HNL}. In an alternating sequence, a KrF excimer laser ($\lambda$ = 248 nm) ablated stoichiometric targets of LaFeO$_3$ and LaCrO$_3$ to produce SL with a 1/1 stacking periodicity. The alternating pattern of LaFeO$_3$ and LaCrO$_3$ unit cells was repeated fifty times, which gives a total thickness of $\sim$ 23 nm. During deposition, the substrate temperature was held at 720 $^{\circ}$C, and an oxygen partial pressure of 10 mTorr was maintained inside the chamber. 

The structural quality of these SL was verified by conventional X-ray Diffraction (XRD). Figure 2 displays the $\theta$-2$\theta$ scan for the SL. As expected, the SL possess a single-phase [111] orientation, which testifies to the quality of the epitaxial growth.  Calculations of the (111) lattice spacing from the XRD data give a value of 2.284 $\AA$, which is 1$\%$ to 2$\%$ larger than in bulk LaCrO$_3$ and LaFeO$_3$ respectively. The anticipated enlargement of the unit cell normal to the surface reflects the lateral compressive strain from the lattice mismatch between STO and the SL. Notice, this lattice spacing is also in agreement with previously reported results \cite{Ueda}. At the same time, the diffraction experiment failed to resolve superstructure peaks indicative of B site cation ordering. These superstructure reflections stem from the two-fold doubling of the perovskite unit cell and are conventionally used to quantify the degree of ordering in rock salt type double perovskites \cite{Garcia-Hernandez}. Our simulations of the XRD pattern for perfect ordering show that predicted intensities of superstructure peaks are two to three orders of magnitude smaller than the principle peaks and therefore below the detection limit of our instrument \cite{scatteringfactors}.

\begin{figure}[t]\vspace{-0pt}
\includegraphics[width=8.5cm]{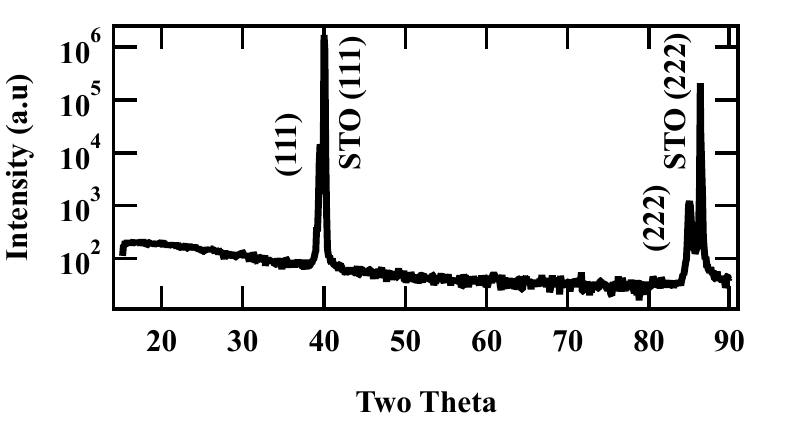}
\caption{\label{XRD111} XRD $\theta$-2$\theta$ scan of a SL.}
\end{figure}

In order to determine the electronic and magnetic structure of the SL, XAS and XMCD experiments were carried out at the 4-ID-C beamline of the Advanced Photon Source (APS) in Argonne National Laboratory. We used circularly polarized soft x-rays at the L-absorption edges of Cr and Fe. Energy resolution was 0.1 eV across the L$_{3, 2}$ edges. Spectra were recorded in total electron yield (TEY) mode on the Fe and Cr L edges at an incident angle of 10$^o$ out of plane. The sample temperature was held at 10 K, and an external magnetic field of 0.1 and 5 Tesla was applied.

Figure 3 compares the Cr and Fe L$_{3, 2}$ XAS spectra of the SL to that of Cr$^{3+}$ (Cr$_2$O$_3$) and Fe$^{3+}$ ($\alpha$-Fe$_2$O$_3$) standards \cite{Theil, Kim}. The Cr/Fe L edge begins around 570/700 eV and extends to 590/730 eV. Consider first the case of Cr. Based on the analogous local electronic structure (Cr in an octahedral cage of six oxygen anions) and strong resemblances in spectral features, the spectrum for the SL can be interrupted based on calculated spectra of Cr$_2$O$_3$. The ground state is high spin t$_{2g}$$^3$ with a 3+ valence. Cr$_2$O$_3$ has a  crystal field splitting parameter 10Dq = 2.0 eV and a small distortion from octahedral symmetry \cite{Theil}. In the same manner, the Fe L edges of the SL and $\alpha$-Fe$_2$O$_3$ can be compared. Fe ions are in the high spin state t$_{2g}$$^3$e$_g$$^2$ and have a valency of 3+. The crystal field splitting parameter  for $\alpha$-Fe$_2$O$_3$ is 10Dq = 1.45 eV, and the ligand field symmetry is O$_h$ \cite{Kuiper}.

\begin{figure}[t]\vspace{-0pt}
\includegraphics[width=8.5cm]{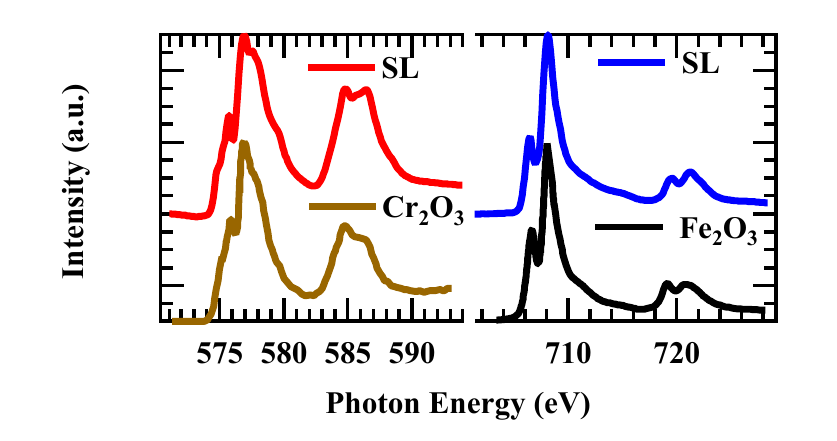}
\caption{\label{XAS111} (Left) XAS at the Cr L edge in red and Cr$_2$O$_3$ \cite{Theil} in green. (Right) XAS at the Fe L edge in blue and Fe$_2$O$_3$ \cite{Kuiper} in black.}
\end{figure}

In Fig. 4, the XMCD spectra on the Cr and Fe L$_{3, 2}$ edges at a sample temperature of 10 K and an external magnetic field of 5 Tesla are presented. The XMCD intensities have been normalized by the XAS peaks. Comparing both the Cr and Fe 2p XMCD, the agreement between the polarities of the two spectra indicates the parallel alignment of magnetic moments on Cr and Fe ions in the presence of magnetic field. Both Cr and Fe show a maximum asymmetry of $\sim$ -2.6$\%$ at the L$_3$ edges. This is unexpected, since for a typical ferromagnet with an average magnetic moment of 3 to 5 $\mu$$_B$ per transition metal site, the anticipated XMCD is 30-40$\%$. The magnitude of the magnetic moments can be estimated by weighting the size of the measured XMCD against standards of known moments with the same valence, ligand field symmetry, and spectral features \cite{Kimura, Kim}. Following this procedure, assignments of 0.31 $\mu$$_B$/Cr and 0.21 $\mu$$_B$/Fe are made for the SL in a magnetic field of 5 T and, when the magnetic field was turned down to 0.1 T, there was no measurable XMCD signal at either edge indicating that the moment at remanence is quite small. For perfect ferromagnetic ordering theoretical values of 3 $\mu$$_B$/Cr and 5 $\mu$$_B$/Fe were expected.  The moments at 5 T are an order of magnitude smaller than Ueda et al, however in thin film experiments caution is required in interpreting magnetometry data, where extrinsic factors can lead to spurious results \cite{Garcia}. It was also noted that due to a conversion error actual magnetization was 0.009 $\mu$$_B$ per site in a measurement at 0.1 T\cite{Meijer}, which is at the level of the noise in our measurements.

\begin{figure}[t]\vspace{-0pt}
\includegraphics[width=8.5cm]{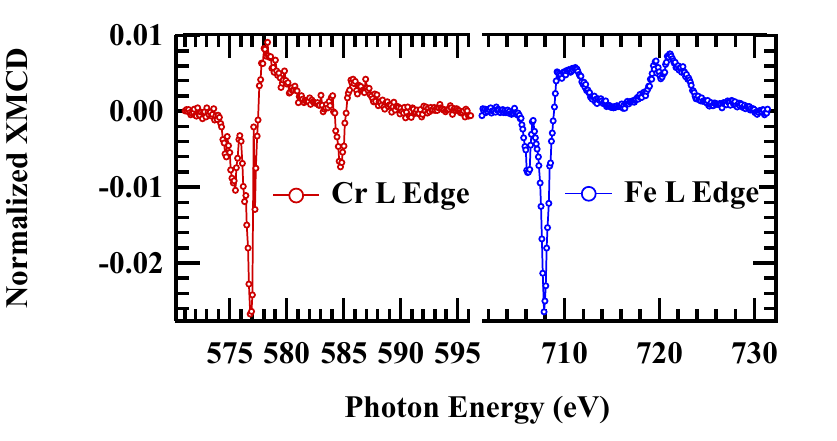}
\caption{\label{XMCD111} XMCD spectrum. (Right) Cr L edge. (Left) Fe L edge at T = 10 K and H = 5 T.}
\end{figure}

There are several possible explanations for the low moment. One is that the films are disordered and not in a rock salt perovskite structure. The complementary ionic radii of trivalent Cr and Fe hinder efforts to produce an order phase leading to antisite disorder, where the perfect alternating order of Fe and Cr is broken. Then the system would would consist of  patches of G-type antiferromagnetic LaFeO$_3$ and LaCrO$_3$, which would have no moment at zero field. However, our experience with G-type anti-ferromagnets is that there moments are hard to cant in this kind of ordered state. To test this we looked at a single layer LaFeO$_3$ film and found no observable XMCD at a field of 5 T and temperature of 10 K. In addition, the difference in lattice parameters for LaFeO$_3$ and LaCrO$_3$ would lead to a distinctly different average c-axis lattice parameter. A second option is to look at other magnetic configurations which have local FM coupling of Cr and Fe but an overall AFM state.  Band-structure calculations have concluded that the ground state might be ferrimagnetic\cite{Pickett1,Miura}. Furthermore, the Goodenough-Kanamori rule are ambiguous for a d$^3$-d$^5$ superexchange. The pd$\sigma$ hybridization between the e$_g$ orbitals of the TM ions and the oxygen p$\sigma$ orbital is FM, while the pd$\pi$ hybridization with the t$_{2g}$ is AFM \cite{Kanamori}. The amount of orbital overlap is greater in the case of pd$\sigma$ hybridization, however there are more paths in pd$\pi$ hybridization. From these ideas and our findings, we can only conclude that the ground state is consistent with canted AFM order.

In summary, we have investigated the local electronic and magnetic environment of Cr and Fe in artificial La$_2$FeCrO$_6$ perovskite by XAS and XMCD. XAS confirms that Cr and Fe are trivalent and in the high spin state. The marginal magnetic moments of Cr and Fe which only appear in large external magnetic fields are consistent with a canted AFM system.

The authors acknowledge fruitful discussions with W. Pickett, V. Pardo, and D. Khomskii. J.C. was supported by DOD-ARO under the Contract No. 0402-17291 and NSF Contract No. DMR-0747808. Work at the Advanced Photon Source, Argonne is supported by the U.S. Department of Energy, Office of Science under grant No. DEAC02-06CH11357. The synthesis work at Oak Ridge National Laboratory (H.N.L.) was sponsored by the Division of Materials Sciences and Engineering, U.S. Department of Energy.

\newpage

\end{document}